\documentclass[fleqn,usenatbib]{mnras}

\usepackage{newtxtext,newtxmath}
\usepackage[T1]{fontenc}
\usepackage{ae,aecompl}
\usepackage{textgreek}
\usepackage{gensymb}
\usepackage{graphicx}	% Including figure files
\usepackage{amsmath}	% Advanced maths commands
\usepackage{amssymb}	% Extra maths symbols

\title[EPIC Simulations of Neptune's Dark Spots]{EPIC Simulations of Neptune's Dark Spots Using an Active Cloud Microphysical Model}

\author[N. Hadland et al.]{Nathan Hadland,$^{1}$\thanks{E-mail: nhadland2016@my.fit.edu (NKH)}
Ramanakumar Sankar,$^{1}$
Raymond Paul LeBeau Jr.,$^{2}$\newauthor
Csaba Palotai$^{1}$
\\
$^{1}$Department of Aerospace, Physics, and Space Sciences, Florida Institute of Technology, Melbourne, FL 32901, USA\\
$^{2}$School of Engineering: Aerospace Engineering, Saint Louis University, Saint Louis, MO 63103, USA\\
}

\date{Accepted XXX. Received YYY; in original form ZZZ}

\pubyear{2020}

\begin{document}
\label{firstpage}
\pagerange{\pageref{firstpage}--\pageref{lastpage}}
\maketitle

% Abstract of the paper
\begin{abstract}
The Great Dark Spot (GDS-89) observed by {\it Voyager 2} was the first of several large-scale vortices observed on Neptune, the most recent of which was observed in 2018 in the northern hemisphere (NDS-2018). Ongoing observations of these features are constraining cloud formation, drift, shape oscillations, and other dynamic properties. In order to effectively model these characteristics, an explicit calculation of methane cloud microphysics is needed. Using an updated version of the Explicit Planetary Isentropic Coordinate General Circulation Model (EPIC GCM) and its active cloud microphysics module to account for the condensation of methane, we investigate the evolution of large scale vortices on Neptune. We model the effect of methane deep abundance and cloud formation on vortex stability and dynamics. In our simulations, the vortex shows a sharp contrast in methane vapor density inside compared to outside the vortex. Methane vapor column density is analogous to optical depth and provides a more consistent tracer to track the vortex, so we use that variable over potential vorticity. We match the meridional drift rate of the GDS and gain an initial insight into the evolution of vortices in the northern hemisphere, such as the NDS-2018. 
\end{abstract}

% Select between one and six entries from the list of approved keywords.
% Don't make up new ones.

%%%%%%%%%%%%%%%%%%%%%%%%%%%%%%%%%%%%%%%%%%%%%%%%%%

%%%%%%%%%%%%%%%%% BODY OF PAPER %%%%%%%%%%%%%%%%%%

\section{Introduction}
Starting with the original Great Dark Spot (GDS-89) observed by {\it Voyager 2}, roughly a half-dozen large geophysical vortices have been observed on the Ice Giants. In 2015-2017, a feature was observed on Neptune in the southern hemisphere (SDS-2015) \citep{Wong2016,Wong2018}. A recent observation as part of the Hubble Outer Planet Atmosphere Legacy (OPAL) program revealed a new dark spot with bright companion clouds in Neptune's northern hemisphere, NDS-2018 \citep{Simon2019}. The structure is the most recent of large scale geophysical features
%vortices 
to be observed on the ice giants. Although these Dark Spots have similar features to large vortices on Jupiter, such as the Great Red Spot, they exhibit dynamical motions such as shape oscillations and latitudinal drift. Neptune anticyclones evolve on a time scale of months \citep{hsu2019} as opposed to large vortices on Jupiter, which evolve over many decades \citep{Ingersoll2004}.  During the {\it Voyager 2} encounter with Neptune beginning in January of 1989, GDS-89 was drifting towards the equator at an approximate rate of 1.3\degree /month \citep{Sromovsky2002}. Post-{\it Voyager 2} observations using the Hubble Space Telescope (HST) revealed that the vortex had dissipated as it approached the equator \citep{HammelLockwood1997}. Many of the observed characteristics have been matched previously by numerical models
(e.g., shape oscillation by \citet{Polvani1990}, drift rate by \citet{LeBeauDowling_Neptune} and companion cloud formation by \citet{Stratman2001}). However, reproducing all of these features simultaneously has remained elusive.

These vortices exhibit surprising variability in terms of evolution, shape, drift, cloud distribution, and shape oscillations, so an explicit calculation of the environmental parameters is required for each vortex observed. However, this is beneficial because more diagnostic information can be obtained from each case than if the spot occurrences were repetitive in terms of their characteristics. For example, in the case of GDS-89, equatorial drift was observed \citep{Smith1989,LeBeauDowling_Neptune} as opposed to the poleward drift of dark spot SDS-2015 \citep{Wong2016,Wong2018}. In addition, GDS-89 had an accompanying cloud feature external to the dark spot, presumably condensed methane, whereas some vortices have centered bright companions such as D2 \citep{Smith1989} and SDS-2015 \citep{Wong2018}. 

\citet{Stratman2001} demonstrated that orographic upwelling could be the cause of companion clouds. An analysis of these overlying orographic cloud features shows promising insight into the dynamics of clouds on Neptune, as the features are directly impacted by the underlying atmospheric structure and the deep methane abundance. To that end, it is necessary to apply a cloud microphysical model to make a more complete representation of dark spots. We use an updated microphysics calculation implemented in the Explicit Planetary Isentropic Coordinate General Circulation Model (EPIC GCM) \citep{Dowling_EPICmodel1998,Dowling2006,Palotai2008} to account for methane cloud microphysics (see Section~\ref{clouds}). We investigate the dynamics of vapor and subsequent persistent cloud formation on Neptune by modelling large scale vortices. 
 
Clouds have been modelled explicitly in recent works to study the effect of convection \citep[e.g.,][]{Sugiyama2011,Li2019} and the GRS \citep{Palotai2014} on Jupiter, but not for Neptune.

We seek to answer the following questions: 
\begin{enumerate}
    \item How does the addition of a methane cloud microphysical model affect the evolution of the vortex? 
    \item How do various methane deep abundance (mole fraction) values affect the vortex? 
\end{enumerate}

To investigate these questions, we use an increased horizontal grid resolution and increased number of vertical layers in the column compared to previous studies. 
Additionally, we use methane vapor to track the vortex drift rate and shape oscillations whereas in previous studies, potential vorticity (PV) was used as the primary parameter to determine the features of the vortex.

\begin{figure}
    \includegraphics[width=\columnwidth]{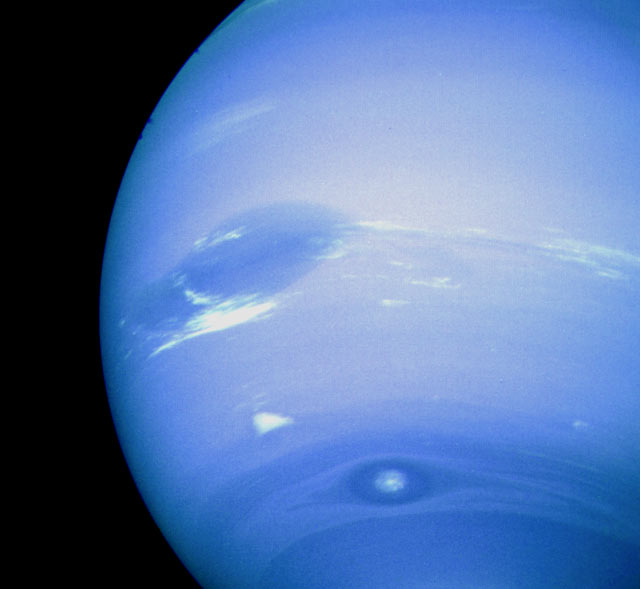}
    \caption{An image of the GDS-89 (dark feature near the center) taken by {\it Voyager 2} in 1989. The high altitude white companion clouds around the GDS are visible. To the South, the Scooter, which is another dark spot observed by {\it Voyager 2} is visible with its distinct centered cloud. Courtesy NASA/JPL-Caltech.}
    \label{fig:GDS-89}
\end{figure}

\section{Methods}
\subsection{EPIC model}
To investigate persistent cloud formation on Neptune, we use the EPIC-GCM with an active hydrological cycle for methane.
EPIC uses a hybrid vertical coordinate, $\zeta$, described by \citet{Dowling2006}. The top of the model uses potential temperature, $\theta$, as a vertical coordinate while the bottom uses a scaled pressure variable, $\sigma$. The transition between the two functions occurs at 10 hPa (see Figure \ref{fig:TP}). Indeed, this transition allows for the use of EPIC on planets without a solid, terrestrial surface (i.e. gas giants) and increased vertical resolution at deeper layers where $\theta$ is nearly constant. 

\subsection{Cloud microphysics} \label{clouds}
We use the active cloud microphysical model from \cite{Palotai2008} which incorporates the condensation of methane in the EPIC model, with the revisions to precipitation given in \cite{Palotai2016DPS}. 

This scheme deals with bulk mass transfer between five explicit phases: vapor, cloud ice, liquid cloud droplet, snow and rain. The last two correspond to precipitation in the model and are subject to sedimentation at the terminal velocity.

Cloud particle sizes are diagnosed using the Gunn-Marshall distribution \citep{GunnMarshall} for snow and the Marshall-Palmer distribution \citep{MarshallPalmer} for rain. Both are log-normal with two free parameters describing the mean and standard-deviation of the particle-size distribution. We apply these distributions due to the low number of parameters (2) and the computational efficiency they provide compared to other, more recently retrieved parameterizations. Furthermore, most Earth-based literature use empirically derived distributions which are accurate for localised conditions, making them a poor choice for porting to gas-giant atmospheres \citep{Palotai2008}.

\subsection{Fall Speed of Particles}

To speed up computation, we parameterize the terminal velocity snow particles of different diameters ($D$) using a power law:
\begin{equation}\label{eq:fall_speed}
    V_t = x D^y \left( \dfrac{p_0}{p}\right)^{\gamma}
\end{equation}
where $x$, $y$ and $\gamma$ are determined by fitting the terminal velocity calculated using  theoretical hydrodynamical principles  \citep{PruppacherKlettBookFallSpeed}. $p_0=1$ bar is the reference pressure. Similarly to \citet{Palotai2008}, we assume that snow is graupellike with a hexagonal shape and follow their formulation for calculating sedimentation velocity as a function of particle size. 

\begin{figure}
    \centering
    \includegraphics[width=\columnwidth]{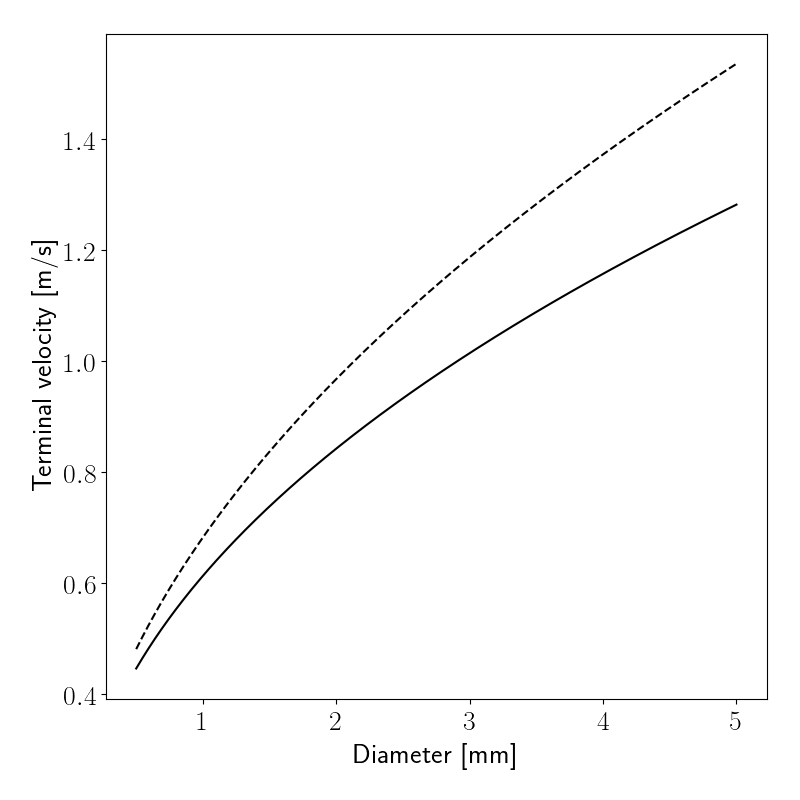}
    \caption{Snow terminal velocity fits for CH$_4$ (solid) and H$_2$S (dashed) at 1 bar on Neptune.}
    \label{fig:termvel}
\end{figure}

A plot of the terminal velocity of CH$_4$ and H$_2$S snow at a pressure of 1 bar is shown in Figure~\ref{fig:termvel}, and the fit parameters for CH$_4$ and H$_2$S on Neptune are given in Table \ref{tab:fall_speed}.

Using the particle size distributions, the net sedimentation rate for a grid cell is calculated by weighting the terminal velocities by the mass of particles in each radius bin. Currently, only snow particles undergo sedimentation since the terminal velocity for cloud ice on Neptune is on the order of a few $\mu$m/s and thus fall only a few meters over the duration of our simulations.

\begin{table}
\begin{tabular}{|c|c|c|}
\hline
\centering
  & CH$_4$ snow & H$_2$S snow  \\ \hline \hline
$x$         & 14.5   & 22.2    \\ \hline
$y$         & 0.458  & 0.504   \\ \hline
$\gamma$    & 0.320  & 0.320   \\ \hline
\end{tabular}
\caption{These constants are determined from Eq. \ref{eq:fall_speed} when $D$ is in meters and $V_t$ is in m/s.}
\label{tab:fall_speed}
\end{table}

\subsection{Model setup} \label{Environmental Initialization}

\begin{figure}
    \includegraphics[width=\columnwidth]{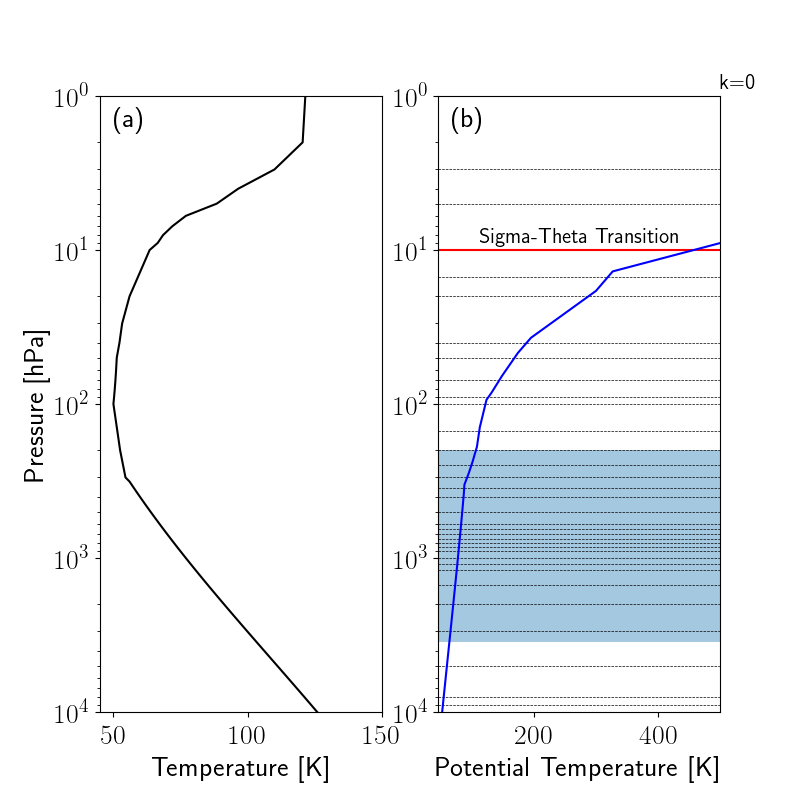}
    \caption{(a) shows the temperature and pressure profile of Neptune used in the EPIC model. (b) shows the potential temperature profile ($\theta$) of Neptune (shown in blue). The horizontal black lines indicate the position of vertical layers used in the model. The GDS-89 extends from approximately 200 hPa to  3.5 bars (shown in the shaded region). The sigma-theta transition is also shown in red at 10 hPa.}
    \label{fig:TP}
\end{figure}

We run 3-dimensional simulations that span $-90\degree$ to $0\degree$ latitude and $-120\degree$ to $120\degree$ longitude with 256 points each resulting in horizontal boxes that are $0.5\degree\times1\degree$ (lat$\times$lon). These limits are sufficient to minimize the effects of the lateral boundaries on the simulated vortex and its associated clouds and prevent the vortex from interacting with itself \citep{LeBeauDowling_Neptune}. The model covers 35 unequally spaced vertical layers between $1$ hPa and $14$ bar (see Figure \ref{fig:TP}), which provides significantly higher resolution than previous studies of the GDS.

\begin{figure}
    \includegraphics[width=\columnwidth]{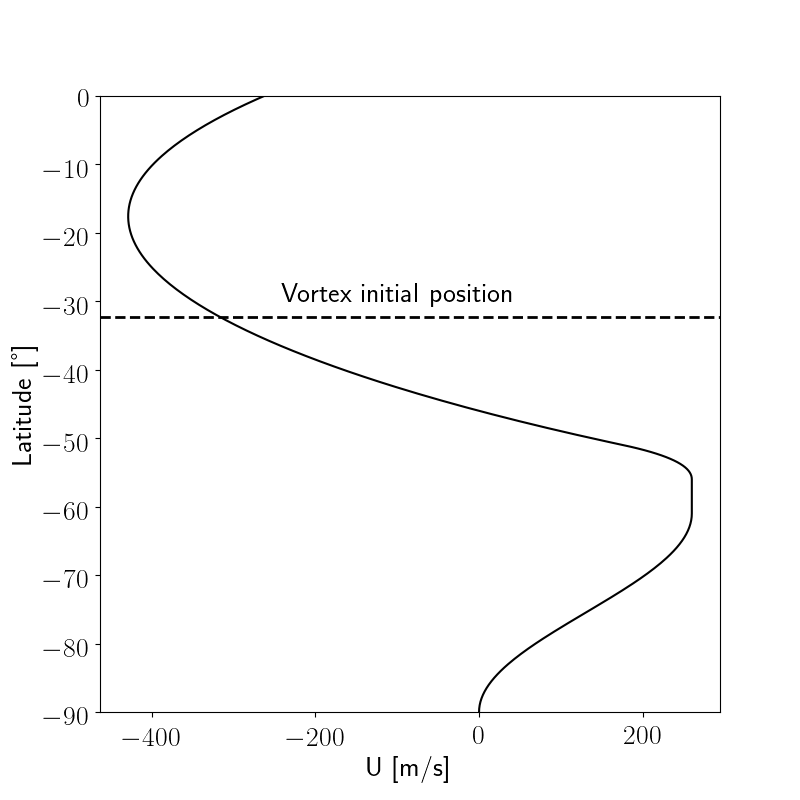}
    \caption{Mean zonal wind profile as a function of latitude using the pseudo PV gradient, $Q_y=\frac{1}{3}$. The vortex is induced at roughly $32\degree$ S latitude. }
    \label{fig:zonal_wind}
\end{figure}

The pressure-temperature profile used in this study is shown in Figure \ref{fig:TP}, which is an idealized curve taken from \citet{LeBeauDowling_Neptune}. The black curve is from previous refinements of {\it Voyager 2} radio-occultation T(P) data from \citet{Conrath1991} and \citet{Stratman2001} took helium at a 19\% mole fraction. This profile is used as the initial input into the model at the equator. The temperature profile throughout the rest of the model is calculated using the thermal wind equation. The layers, shown on Figure \ref{fig:TP} as horizontal lines on the right, were chosen in order to increase the resolution around the initial location of the spot (approximately 1 bar) in the vertical column and the methane cloud deck.  
In addition, since the model reaches up to 1 hPa, we can set the $\sigma-\theta$
transition at 10 hPa, or between $k = 2$ and $k = 3$ indicated by the intersection of the horizontal red line and the potential temperature profile in blue. We initialize equilibrium simulation using the idealized zonal wind profile $Q_y = \frac{1}{3}$ from \citet{LeBeauDowling_Neptune}
, which closely matches the drift rate of the GDS-89 as opposed to other idealized curves (see Fig. \ref{fig:zonal_wind} and Section \ref{Drift_rate}). We run 130 day simulations to investigate stability, drift rate, and companion clouds throughout the model, with the ultimate goal of achieving similarities to the observations of GDS-89 and other vortices. Given that the observed periodicity in the oscillation of the GDS-89 was about 8 days \citep{Sromovsky1993}, the model should be able to capture several periods of wobble in the vortex.

Initially, the geopotential heights are determined in the model by integrating the hydrostatic equation. Due to numerical errors, the model is initially slightly unstable, so we run the model for 5 days to let the winds stabilize. During this time, cloud microphysics is turned off so as to not have spurious cloud growth. The spot is added to the model at the end of this equilibrium phase. 

\subsection{Methane abundance}
The primary goal of this work is to analyse the effect of methane on the dynamics and stability of the GDS-89. To that end, we vary the deep abundance (mole fraction) of carbon and initial ambient humidity in our test cases. 

In the nominal case, we take the standard value of $40\times$ the solar [C/H] fraction for Neptune \citep{Baines1995}, or an approximate CH$_4$ mole fraction of $\textit{f}_{\text{CH}_4} = 0.022^{+0.005}_{-0.006}$. The vertical methane abundance profile is initialized by a ``cold-trap" model, where the mole fraction at the deepest layer is defined by a constant value (the deep abundance). Successive layers are limited by the minimum of the saturation mole fraction and the mixing ratio of the layer below. In our model, we take the initial relative humidity for all ``wet" cases to be 95\% to prevent spurious cloud growth in the first timestep. In a 3-dimensional model, advection of methane vapor and cooling produce localized supersaturation, which leads to cloud formation.

We run 5 cases which are shown in Table~\ref{Cases}. Three include active cloud microphysics at different methane deep abundances. In the passive case, methane vapor is added to the model but does not condense. We also include a dry (H/He atmosphere) case. These different parameters test both the effect of the additional mass from methane vapor and the latent heat release from cloud formation on the dynamics of the GDS. 

\begin{table}
\centering
\begin{tabular}{ccc}
\hline

Case    & Cloud microphysics & [C/H] relative to solar \\ \hline\hline
1       & On                 & $20$                                    \\ \hline
2       & On                 & $40$                                    \\ \hline
3       & On                 & $80$                                    \\ \hline
4       & Off                & $40$                                    \\ \hline
5       & Off                & -                                       \\ \hline
\end{tabular}
\caption{A summary of cases tested in this study. For case 4, methane was added but cloud microphysical processes were turned off (i.e. methane vapor was subject only to advection). Case 5 is a dry model which has no added methane.}
\label{Cases}
\end{table}

\subsection{Addition of the Vortex} \label{sec:addvort}

We utilize the vortex initialization as detailed in \citet{LeBeauDowling_Neptune}, who use a Kida vortex model.
The vortex begins at a latitude of $32\degree$ S at a pressure level of 1000 hPa ($k = 24$). The vortex extends roughly a scale height vertically in both directions (see Fig. \ref{fig:TP}). The initial vortex adjustment period is about $6-8$ model days during which the shape of the spot changes drastically. As such, we begin our analysis after this adjustment period; $10$ days into the model. 

The induction of the vortex creates a region of anomalous potential vorticity (PV) compared to the ambient zonal wind shear, defined as the difference in PV between the test case and the corresponding equilibrium simulation. PV has been used in previous modelling works \citep[e.g.][]{LeBeauDowling_Neptune, Stratman2001} to study the GDS and is a useful tracer in dry models to track the vortex. We can define the extent of the GDS by a contour of constant anomalous PV at a pressure level of $1$ bar or $500$ hPa (the centre of initialized vortex). The shape and location of the vortex are then quantified by fitting an ellipse to these contours using the method of \citet{HalirFlusser1998}.

However, the PV of the simulated vortex is difficult to compare with {\it Voyager 2} observations due to the lack of precise wind speed measurements. The change in the atmospheric structure by the addition of the vortex leads to a distinct variation in methane abundance inside the GDS, which can be used to track the dynamics of the spot in a similar way to PV and can also be compared with the observed spot opacity. By integrating the methane vapor density with depth, we obtain the column density (CD), which is analogous to optical depth in the upper troposphere. We then take the ratio of the CD to the equilibrium (see Section \ref{Environmental Initialization}) and proceed similarly, by fitting an ellipse to a contour of constant CD. Hereafter, references to CD implies this ratio, unless otherwise specified.

\section{Results}

\subsection{1D Simulations}
We first run a 1-dimensional case with methane cloud formation enabled to test the microphysical processes and input parameters. We initialize the atmosphere with 150 layers between 1 hPa and 14 bar. We use a nominal value of $40\times$ the solar [C/H] fraction, and the model is initially saturated at 100\% relative humidity everywhere. 

\begin{figure}
    \includegraphics[width=\columnwidth]{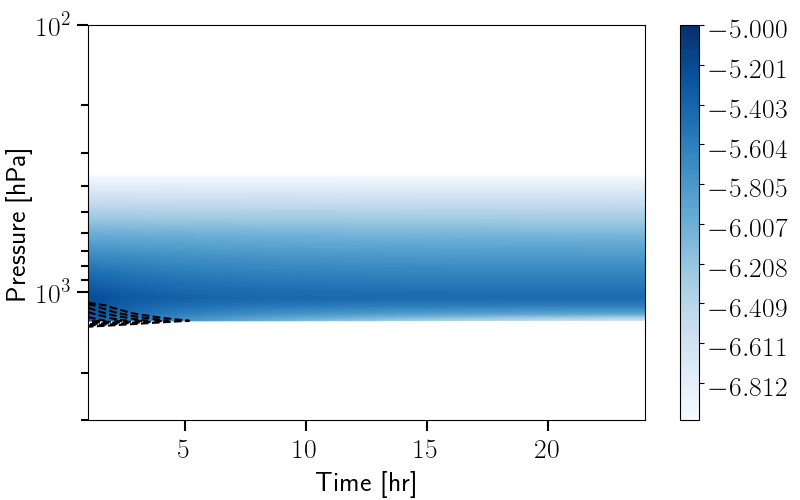}
    \caption{Results from 1D simulations. Methane ice cloud is plotted in light blue and snow is plotted in black dashed contours. Colorbar corresponds to the base-10 log of the cloud density in kg/m$^3$. }
    \label{fig:1dsim}
\end{figure}

From equilibrium cloud condensation models (ECCM) \citep[e.g.,][]{AtreyaWong2005}, methane ice clouds are predicted to form at around 15 bar. Figure~\ref{fig:1dsim} shows the cloud as a function of time. The base of the clouds in the model are at about 1050 hPa. Snow is shown with the black contours and is precipitated out in a few hours. Cloud ice is most dense near the base and has a typical density of $\sim 10^{-5}$ kg/m$^3$.

\subsection{Potential vorticity} \label{vorticity_dynamics}
\begin{figure}
    \includegraphics[width=\columnwidth]{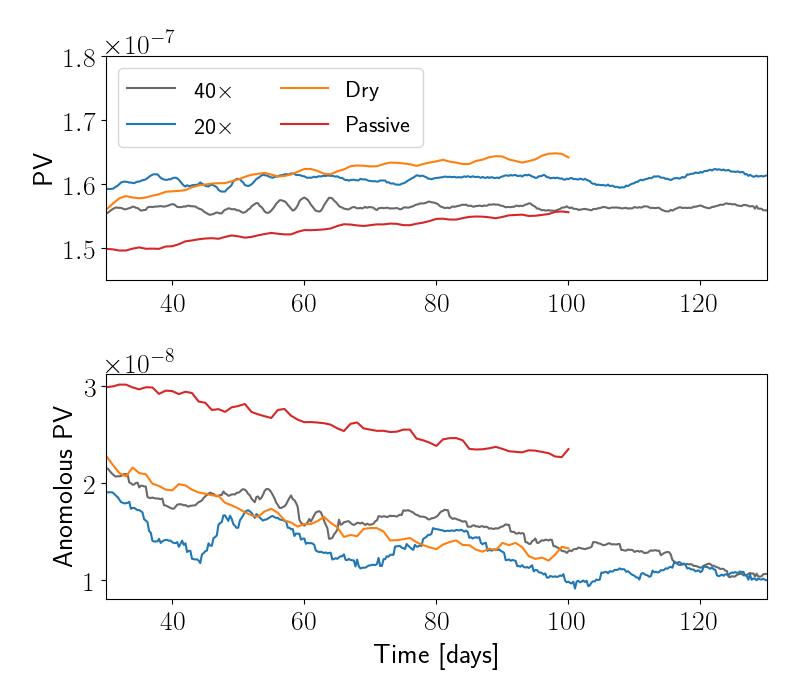}
    \caption{Temporal variation PV and anomalous PV %(from the equilibrium) 
    (both in units of m$^2$ K s$^{-1}$ kg$^{-1}$)
    at 500 hPa inside the simulated GDS in our test cases using CD to track the vortex. In the dry case, only anomalous PV at 500 hPa was used to track the simulated GDS due to the lack of methane. } 
    \label{fig:pv_evolution}
\end{figure}

To fit the vortex, we use the anomalous potential vorticity as a tracer. We also tracked the evolution of the anomalous PV within the vortex, as shown in Figure~\ref{fig:pv_evolution}. As the vortex drifted equatorward, the background wind shear and the background potential vorticity increased. Consequently, although the total PV of the vortex itself was conserved, the anomalous potential vorticity of the vortex decreased with time, making it difficult to consistently track the vortex. Thus, we are limited to only the first 100 days of tracking the vortex with this method, after which the anomalous PV of the vortex was comparable to ambient noise.

\begin{figure*}
    \centering
    \includegraphics[width=1\textwidth]{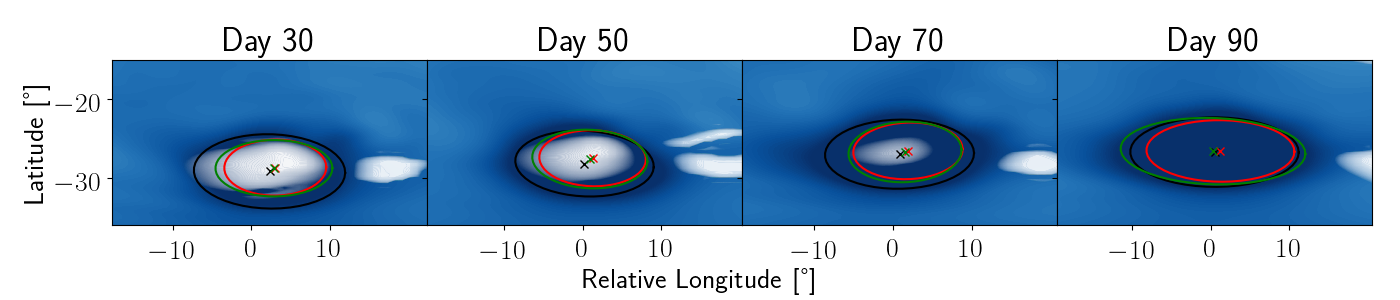}
    \caption{Snapshots of the $40\times$ solar methane abundance simulation showing CD fit (black), PV fit at 1 bar (red), and PV fit at 500 hPa (green) to the vortex, along with their respective centres ($\times$). Methane column density is plotted in the background in blue, with the darker regions corresponding to lower CD (less vapour) with a lower limit of $-0.015$ kg/m$^2$ and an upper limit of $0.4$ kg/m$^2$ . The PV is shifted by about $1.5\degree$ to the north east from the CD centre and decreases over time.}  
    \label{fig:cd_pv}
\end{figure*}

We fit the anomalous PV at two levels: 1000 hPa and 500 hPa (Figure~\ref{fig:cd_pv}), corresponding roughly to the vertical extent of the vortex. Below roughly 2000 hPa, the vortex is very weak and the anomalous PV is almost negligible. We find that the vortex is fairly uniform between the two layers, as the vortex is mostly equal in size and at the same position. The 1000 hPa anomalous PV decreases faster than the 500 hPa layer, and thus for all further analysis we report values from the vortex tracked using the 500 hPa anomalous PV.

\subsection{Drift Rate}\label{Drift_rate}
\citet{LeBeauDowling_Neptune} investigated the equatorial drift of GDS-89 by using constant-vorticity gradient zonal wind profiles (Fig. \ref{fig:zonal_wind}) and found that the value of the mid-latitude pseudo PV gradient, $Q_y$ is the primary influencer on meridionial drift. Using $Q_y=\frac{1}{3}$, our experiments match the observed drift rate of the GDS-89 of $1.3\degree$ per month \citep{Sromovsky1993}.

\begin{figure}
    \includegraphics[width=\columnwidth]{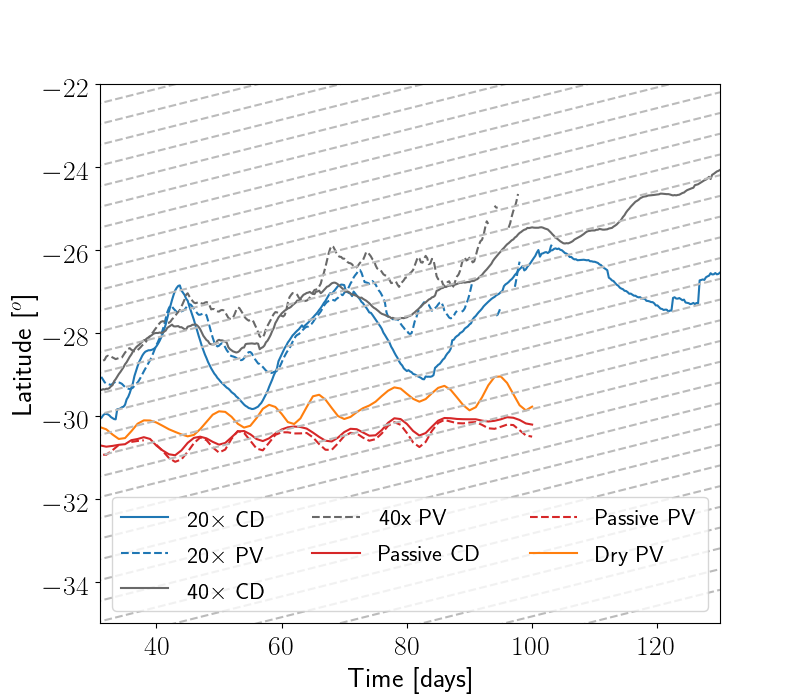}
    \caption{Drift rate of the simulated GDS in our test cases. We used both methane column density (solid) and potential vorticity (dashed) at 500 hPa to fit the ellipse.  In the dry case, only PV at 500 hPa was used to track the simulated GDS due to the lack of methane. he dashed grey lines are the average meridional drift rate of the GDS-89 observed by {\it Voyager 2} \citep{Sromovsky1993}. }
    \label{fig:drift_rate}
\end{figure}

Figure \ref{fig:drift_rate} shows the different cases investigated in this study and their latitudinal drift relative to the approximate observed drift rate of the GDS. Two lines per case are depicted, one for the PV at 500 hPa (dashed) and one for the methane column density (solid). Table \ref{drift_rate_table} summarizes the average drift rate for each case investigated.

Case 2 achieves the closest drift to the observed value of GDS-89, followed by Case 1. The  passive and dry cases (Cases 4 and 5 respectively) have the slowest drift. The $80\times$ (Case 3) struggles to maintain dynamic stability throughout the simulation. We could not fit an ellipse around the vortex since it did not have a uniform shape and it dissipated after approximately 60 days. 

\begin{table}
\centering
\begin{tabular}{ccc}
\hline

Case    & CD Drift Rate [$\degree$/month] & PV Drift Rate [$\degree$/month]\\ \hline\hline
1       & $0.82$   & $0.90$                                                   \\ \hline
2       & $1.36$ & $1.26$                                                     \\ \hline
3       & -        & -                                            \\ \hline
4       & $0.27$ & $0.31$                                                    \\ \hline
5       & - & $0.46$                                                   \\ \hline
\end{tabular}
\caption{A summary of the approximate drift rate using either CD or PV at 500 hPa for each case. For case 3($80\times$), we were unable to generate a fit. Case 5 is a dry model so PV alone was used to track the vortex.}
\label{drift_rate_table}
\end{table}

\subsection{Shape Oscillations}\label{shape_oscillations}
\begin{figure}
    \includegraphics[width=\columnwidth]{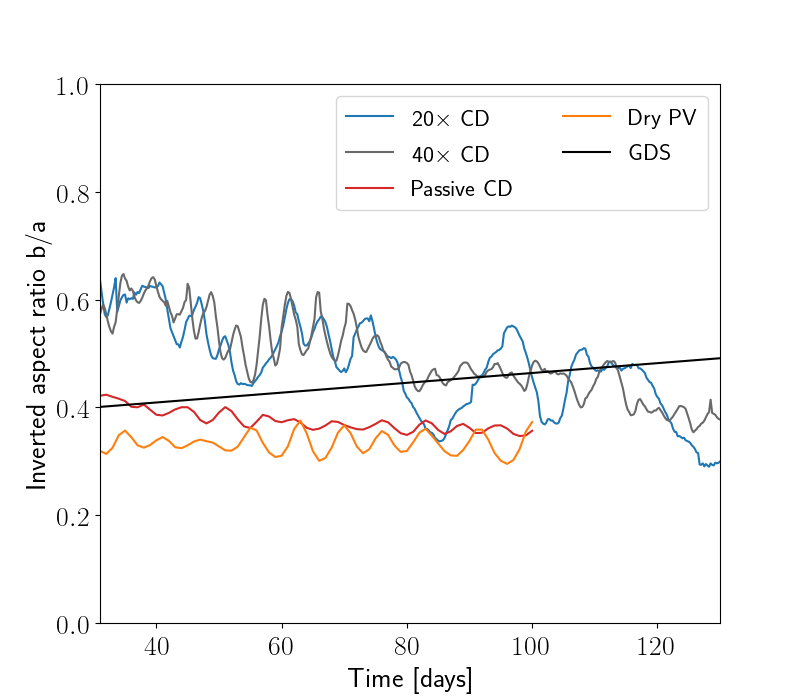}
    \caption{The inverted aspect ratio of the ellipse (semi-major divided by semi-minor) for the different test cases using methane column density to fit the simulated GDS. The solid black line is the average increase in the inverted aspect ratio of the GDS-89 observed by {\it Voyager 2} \citep{Sromovsky1993}. The period nature of the vortex (i.e. the `rolling motion') is evident in all cases, but at different time scales. }
    \label{fig:shape_oscillation}
\end{figure}

Along with meridionial drift, time dependent oscillations in the shape of the vortex are of interest. Figure \ref{fig:shape_oscillation} shows the inverted aspect ratio ($b/a$) of the fitted ellipse for CD (Cases 1-4) and the PV at 1 bar (Case 5). In all cases, we correct for geometric effects by converting the extents to physical lengths using an equatorial radius of $24,760$ km and polar radius of $24,343$ km.  Clearly, these simulations exhibit complex dynamical variability in shape. The passive (Case 4) and dry (Case 5) cases have a lower inverted aspect ratio than the active cases (1-3) at around 0.3 as opposed to 0.4. Additionally, for the active cases, there does not appear to be a predictable frequency of oscillation and a Fourier power spectrum did not reveal any strong periodicity. The semi-major axis of the active cases are increasing without a corresponding increase in the minor axis resulting in a more elliptic vortex with time. Conversely, the passive and dry cases have the opposite occurring (more circular). 

The GDS-89 was observed to increase in its semi-minor axis ($b$) at a rate of $3.8 \times 10^{-5}$ h$^{-1}$ \citep{Sromovsky1993} as it drifted northward, while the semi-major axis ($a$) was roughly constant, resulting a more circular vortex with time. This is in contradiction to our cloudy simulations where we see the vortex getting more elliptic by stretching along the zonal axis as it drifts northward while maintaining a nearly constant meridional height (Figure~\ref{fig:cd_pv}). In the passive and dry case, there is very little change in the aspect ratio throughout the run, and they are much closer to the observed aspect ratio trend observed by {\it Voyager 2}. 

\subsection{Companion Clouds}

\begin{figure}
    \includegraphics[width=\columnwidth]{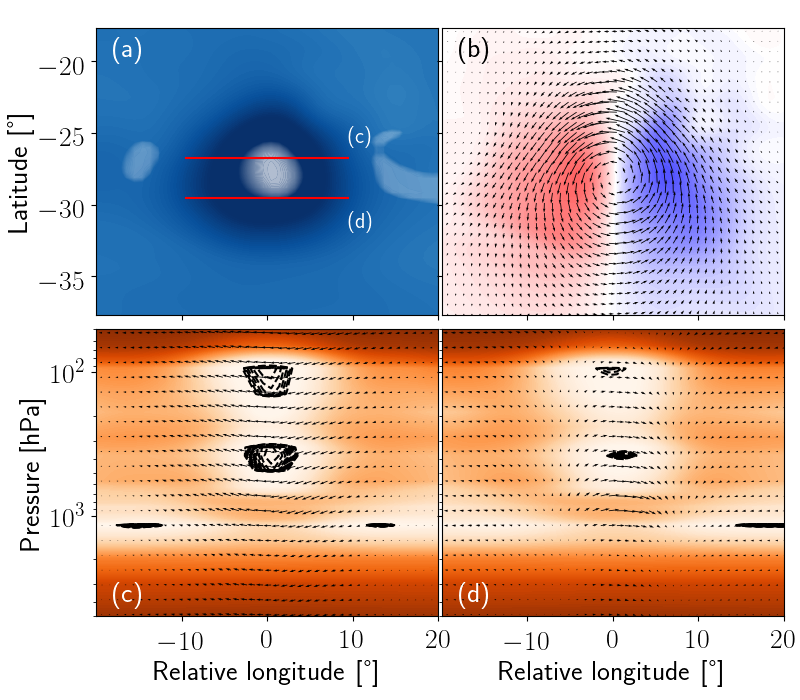}
    \caption{Model output at day 44. (a) shows the top-down view of methane column density (blue) and cloud (white). (b) shows the vertical wind with upwelling in red and downwelling in blue, with wind vectors in the comoving frame. (c) and (d) are longitudinal slices showing methane relative humidity, with brighter corresponding to saturated regions and darker being drier. Wind vectors show the local circulation within the vortex and the black contours are ice cloud densities. The location of the slices are shown in (a) in red. }
    \label{fig:comp_cloud}
\end{figure}

Figure~\ref{fig:comp_cloud} (a) shows a snapshot of the model output with the methane column density in blue and companion clouds in white. There are two distinct types of clouds that interact with the vortex: those that are interior (within the horizontal extents of the vortex) and that are exterior to the vortex (Figure~\ref{fig:comp_cloud} a). The exterior clouds form a thin layer at roughly 1 bar (Fig~\ref{fig:comp_cloud} c, d), while the interior clouds form much higher and are extended vertically.  

The interior clouds form at two discrete levels: one close to the tropopause at $\sim100$ hPa and the other further down around $400-500$ hPa. The upper level cloud particles are small due to being in a thinner environment and do not reach the critical diameter of $500\mu$m to precipitate. The humid environment where the clouds form (bright in Fig~\ref{fig:comp_cloud} c, d) are retained throughout the vortex lifetime due to the thermal structure of the vortex. 

The deeper cloud is denser and precipitates immediately. However, both clouds are long lived and last about 30 days after the initial adjustment period before dissipating. The particle sizes in both clouds, which are diagnosed from the cloud water content \citep{Palotai2008}, are on the order of $300-500\mu$m with the larger particles forming in the lower cloud. These sizes are comparable to cold cirrus clouds on Earth \citep{Heymsfield2017}, which is a similar habit to the interior clouds in the GDS. 

\begin{figure}
    \includegraphics[width=\columnwidth]{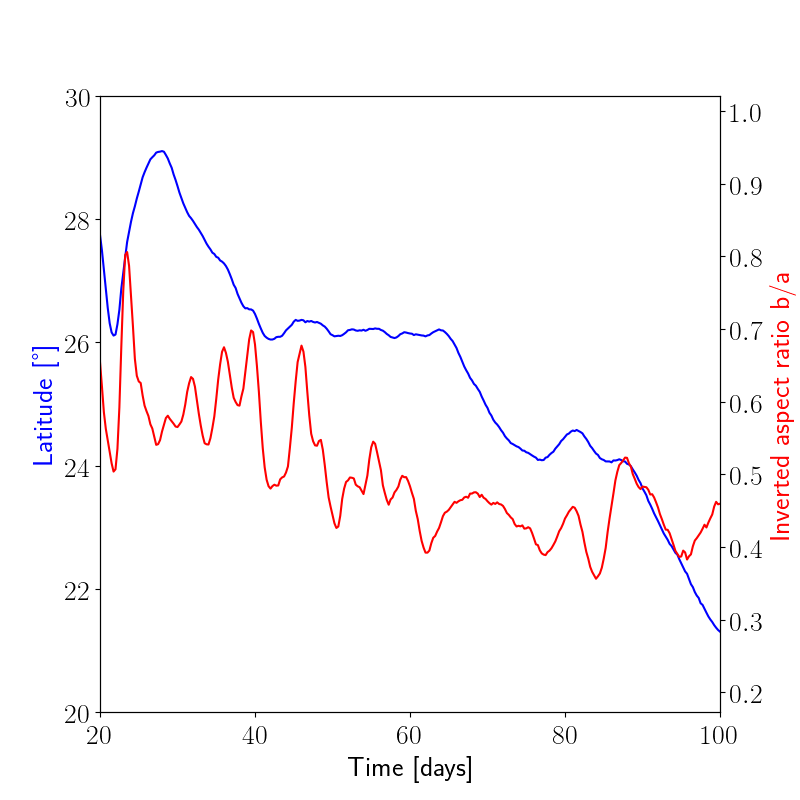}
    \caption{Drift rate (blue) and aspect ratio (red) of NDS-2018 from model output. }
    \label{fig:NDS_drift}
\end{figure}
\subsection{NDS-2018}

\begin{figure*}
    \centering
    \includegraphics[width=1\textwidth]{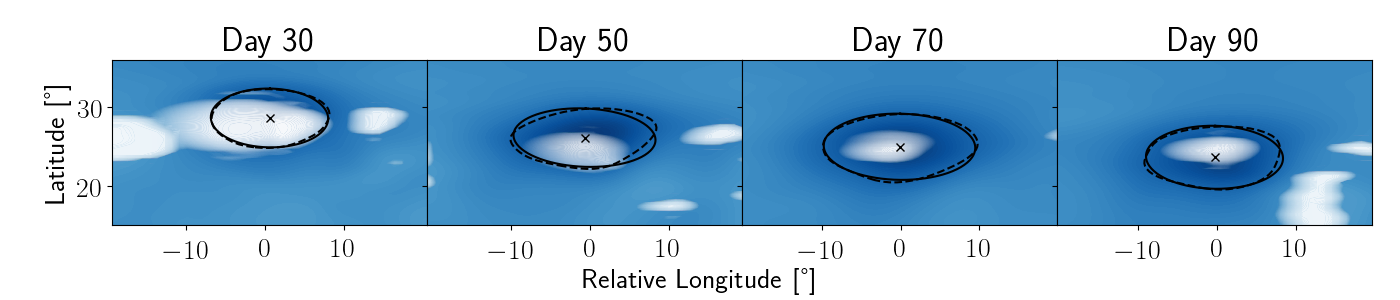}
    \caption{Select timesteps from a 100 day simulation of Neptune's NDS-2018 using the EPIC model with active microphysics for methane and $40\times$ solar methane abundance. The blue is integrated column density of methane vapor with darker regions indicating areas of low density. The solid line is the fitted ellipse using dashed line, which is the vapor field contour.}
    \label{fig:nds}
\end{figure*}

In 2018, Hubble Space Telescope visible wavelength imaging revealed the presence of a dark spot in the northern hemisphere (labelled NDS-2018) \citep{Simon2019}. OPAL observations showed that there was increase in active cloud formation in the regions for 1-2 years leading up to the development of the spot. The NDS-2018 is similar in size and shape to the GDS-89 \citep{Simon2019}. Consequently, in order to constrain the dynamics of this feature, we run simulations with identical vortex initialization, domain, and resolution as the GDS cases. We run 100 day simulations of this feature using an active cloud microphysical model. The vortex was initialized at $32\degree$ N latitude with a 
$40\times$ solar [C/H] fraction and an initial relative humidity of 95\%. Fig. \ref{fig:nds} shows select timesteps of this NDS simulation. 

Due to the symmetrical nature of the zonal wind profile used in this study (Fig \ref{fig:zonal_wind}), equatorial drift was observed in the simulated NDS case similar to the simulated GDS cases (Fig. \ref{fig:NDS_drift}). The average approximate drift rate found in this study for the NDS-2018 was roughly $3.2\degree$ per month. In contrast to the surprisingly consistent latitudinal wobble observed in the southern hemisphere for the GDS cases, the NDS case has inconsistent oscillations in latitude. This simulation also saw a decrease in inverted aspect ratio throughout the 100 day simulation with an average value of around 0.4 consistent with the value of $b/a=0.45$ observed by \citet{Simon2019} for the NDS-2018.

The cloud formation in the NDS simulations were also similar to the GDS cases, with a set of interior and exterior clouds, although the interior clouds were not perfectly centered like the GDS simulations. These are likely a result of the much more irregular dynamics of the NDS vortex in our simulations.

\section{Discussion}
Large scale vortices that have been observed on Neptune exhibit surprising variability in terms of shape, drift, and lifetime. Consequently, they are some of the most dynamic features in the Solar System. Due to these variations, a variety of vortex characteristics can be examined at short timescales. Although we are constrained by limited observational data, recent developments in the accuracy of models using active microphysics can help us investigate the dynamics of these features. 

\subsection{Methane deep abundance} 

The various cases discussed here exhibit a wide range in both the meridionial drift and the dynamic time-dependent variations in shape of the GDS-89. 

We tracked the vortex using both anomalous PV and methane column density (Figure~\ref{fig:cd_pv}), and found that using the CD allows us to track the vortex for much longer due to the anomalous PV decreasing as the vortex moves through different shear environments. These cases differ significantly in the rate of latitudinal drift, the change in aspect ratio and the periodicity of the oscillations. Cases 1 and 2 ($20\times$ and $40\times$ solar methane abundance with active microphysics) are best matches for the drift rate (with Case 2 reproducing the drift almost perfectly), but do not show the observed decrease in ellipticity of the GDS-89. The passive and dry (Cases 4 and 5) are poor matches for the drift rate but are better at reproducing the observed slow change in the aspect ratio. Surprisingly, the $80\times$ simulation failed to achieve a stable feature, and dissipates after 60 days. It should be noted that the high fidelity observations of the GDS started above $\sim-26\degree$ latitude \citep{Sromovsky1993}, which is further north compared to where most of our simulations reach, and it is likely that we were probing an earlier part of the GDS-89's trajectory. A study of starting the vortex further north will need to be done to investigate this. Ultimately, the stability of Cases 1 and 2 and their consistency with {\it Voyager 2} observations reinforces the measurements of a roughly $20-40\times$ solar [C/H] ratio on Neptune. 

\subsection{Cloud microphysics}

Cloud microphysics has a much more drastic effect on the dynamics of the vortex compared to simply the addition of methane, as the passive case behaves similarly to the dry case. It is, however, much more difficult to trace the process(es) responsible for these effects. Due to the low density of the methane clouds, coupled with the low latent heat release of methane, it is unlikely to be just latent heating that drives this difference.

Clouds were present throughout the simulation and interacted strongly with the vortex. Similar to \citet{Stratman2001}, we observed multi-layered clouds both within and without the vortex in our simulations (Figure~\ref{fig:comp_cloud}). Their eastward cloud forms much higher than ours. However, this may be a function of vertical resolution and interpolation between the layers. In general, both results agree with the humid environment within the vortex and a region of dry air directly below. We are unable to reproduce the persistent poleward cloud. 
\citet{Stratman2001} showed that vertical location of the vortex strongly influences the formation of clouds and indeed observed the poleward cloud only in certain configurations. The vertical extent and location of the spot in our simulations is held constant to reduce the number of free parameters, and we will investigate the effect of changing them in future simulations.

The deep cloud within the vortex formed snow and had cloud particles exceeded $300\mu$m in diameter based on the assumption of hexagonal plate-like particles. Using {\it Voyager 2} IRIS spectra, \citet{Conrath1991} obtained particle sizes between $0.3-30\mu$m, assuming Mie scattering by spherical particles. In our model, we assume ice and snow to be hexagonal plates (as determined for Earth clouds) resulting in much larger diameters than the equivalent spheres for the same particle masses. The equivalent radii for the spherical cloud particles in our simulations is about $50-70\mu$m, for the same particle mass (i.e. the ``wetted radius"). \citet{Kinne1989} noted that using spherical ice crystals underestimates the albedo by up to $15\%$ due to greater forward scattering, depending on the shape of the ice particle. Therefore, the discrepancy in cloud particle sizes on Neptune requires further study in the avenue of experimentally determining methane nucleation in a H/He environment and scattering for non-spherical geometries. Both of these would further refine the microphysical parameters in our model to better represent gas-giant atmospheres.

\subsection{Optical depth}

Strong methane absorption in Neptune's atmosphere largely correlates with a higher optical depth \citep{Hueso2017}. In our simulations, we find that the vortex exists as an area of low methane vapor density compared to the surrounding atmosphere. This is due to the modified thermal structure of the atmosphere as a consequence of introducing the vortex (Figure~\ref{fig:cd_pv}). All cases (except for Case 5 - which contains no methane vapor) reciprocate this phenomenon and decreased methane column density persist throughout the simulation. The reduced column density results in a decreased optical depth over the vortex compared to the ambient background atmosphere. We interpret this as being able to see deeper into the atmosphere, producing the ``darkness" of the GDS-89. This is, indeed, an `apples-to-oranges' comparison, and applying a forward radiative transfer (RT) model to simulation outputs would lend additional insight into this discussion. We are working on an RT model that will address this in future studies.

\subsection{Applications to the NDS-2018}

The NDS-2018 simulation has surprising differences from our GDS test cases. There is much more irregularity in the equatorward drift of the vortex and a higher drift rate. The shape oscillation was similar to our wet GDS cases, with the vortex becoming more elliptic with time.

These differences could be due to the {\it Voyager}-era zonal wind profile we use in our models. Recent OPAL observations have provided new zonal wind data and even shown evidence of weak vertical wind shear \citep{Tollefson2018}. \citet{Simon2019} suggest that a zonal wind gradient of $\sim4\times$ that of the \citet{Sromovsky1993} fit is required to match the observed aspect ratio of the NDS-2018, or $du/dy\sim5.4\times10^{-5}$ s$^{-1}$. The Q$_y=1/3$ zonal wind profile we use has a value of $du/dy\sim 3\times10^{-5}$ s$^{-1}$ around $30\degree$ latitude. Testing different zonal wind gradients is not part of this study however, and will be introduced into future sims. Furthermore, without additional observations, it is difficult to verify the accuracy of the drift rate and shape oscillations we measured in the model. It is unknown whether the spot is still present because Hubble observations are limited.

\section{Conclusion}

In this study, we have analysed the effect of methane abundance and cloud formation on the dynamics of the vortex. While the addition of active microphysics improves the meridional drift rate of the vortex, there are other observations which are difficult to reproduce, such as the consistent periodicity of the shape oscillation or the increase in inverse aspect ratio. A further study of the parameter space that describes the spot is required, such as changing the initial vortex location and size (both vertically and horizontally). Additionally, \citet{Tollefson2018} assumes a $\times 2$ or an $\times 4$ depletion of methane at the mid-latitudes and towards the poles and an increase in mixing ratio near the equator based on idealized equations. These recent observations would be beneficial in the study of the newer spots such as the NDS-2018, and indeed are likely required to explain their dynamics. 

The active microphysics package implemented in this study has a clear effect on the dynamical motions of the vortex, such as meridional drift and shape oscillations. Using a $40\times$ or $20\times$ methane deep abundance, we achieve a stable simulation for over 120 days. We also see persistent bright companion cloud formation throughout the simulation. Additionally, the vortex exists as an area of low density which may be correlated to a decrease in optical depth. Finally, we found a possible scenario for the dynamical evolution of the NDS-2018 including its equatorial drift and the vortex becoming more elliptic with time. However, additional observations are needed to constrain these simulations. 

\section*{Acknowledgements}
This research was supported in part by the NASA Solar System Workings Program grant NNX16A203G and the DPS Hartmann Student Travel Grant Program. N.H. thanks the Astronaut Scholarship Foundation for their support. We also thank Noah Nodolski for his help and the anonymous reviewer for their input.

\vspace{-1em}
\section*{Data Availability}
The data underlying this article will be shared on reasonable request to the corresponding author.

%%%%%%%%%%%%%%%%%%%%%%%%%%%%%%%%%%%%%%%%%%%%%%%%%%

%%%%%%%%%%%%%%%%%%%% REFERENCES %%%%%%%%%%%%%%%%%%

% The best way to enter references is to use BibTeX:

%\bibliographystyle{mnras}
%\bibliography{example} % if your bibtex file is called example.bib

% Alternatively you could enter them by hand, like this:
% This method is tedious and prone to error if you have lots of references
\vspace{-2em}
\bibliographystyle{mnras}
\bibliography{hadland2020} % if your bibtex file is called example.bib

% Don't change these lines
\bsp	% typesetting comment
\label{lastpage}
\end{document}